# Building the 'JMMC Stellar Diameters Catalog' using SearchCal

Sylvain Lafrasse*[a], Guillaume Mella[a], Daniel Bonneau[b],
Gilles Duvert[a], Xavier Delfosse[a], Olivier Chesneau[b], Alain Chelli[a]

[a] Université Joseph Fourier - Grenoble 1 / CNRS, Laboratoire d'Astrophysique de Grenoble (LAOG) UMR 5571, BP 53, 38041 Grenoble Cedex 09, France

[b] UMR 6525 H. Fizeau, Univ. Nice Sophia Antipolis, CNRS, Observatoire de la Côte d'Azur, Av. Copernic, 06130 Grasse, France

## ABSTRACT

The JMMC[1] Calibrator Workgroup has long developed methods to ascertain the angular diameter of stars, and provides this expertise in the SearchCal[2] software. SearchCal dynamically finds calibrators near science objects by querying CDS[3] hosted catalogs according to observational parameters. Initially limited to bright objects (K magnitude ≤ 5.5), it has been upgraded with a new method providing calibrators without any magnitude limit but those of queried catalogs. We introduce here a new static catalog of stellar diameters, containing more than 38000 entries, obtained from SearchCal results aggregation on the whole celestial sphere, complete for all stars with HIPPARCOS[4] parallaxes. We detail the methods and tools used to produce and study this catalog, and compare the static catalog approach with the dynamical querying provided by SearchCal engine. We also introduce a new Virtual Observatory service, enabling the reporting of, and querying about, stars flagged as "bad calibrators" by astronomers, adding this ever-growing database to our SearchCal service.

**Keywords:** JMMC, SearchCal, optical interferometry, calibrator, star diameter, virtual observatory

## 1. INTRODUCTION

A good calibration of visibilities is paramount for optical interferometry. This calibration is performed using "calibrators", stars whose angular diameter is ascertained by direct or indirect methods. SearchCal is one piece of the JMMC software portfolio dedicated to OLBI (Optical Long Baseline Interferometry) observation preparation. It aims to compute those diameters with surface brightness relationship[5] using star photometry retrieved from catalogs available at CDS. To overcome latency and dependency on network resources for the frequently used bright star queries, we compiled a new static catalog of bright star diameters, known as JSDC[6] (JMMC Stellar Diameters Catalog), containing 38472 entries. A new version of SearchCal, dedicated to objects fainter than K=5, allows an angular diameters determination without prior knowledge of the parallaxes[7]. But in the present version of JSDC, we impose that the selected stars have a parallax measurement and we use the "Bright" option of SearchCal presented in Bonneau et al. (2006). Such choice limits the number of faint calibrators available but provides greater accuracy in the determination of angular diameters. We will detail how this catalog is built and analyzed. We also present a corollary effort known as BadCal[8], to collect data about stars observed and flagged as "bad calibrators".

*sylvain.lafrasse@obs.ujf-grenoble.fr; phone +33 4 76 63 55 30; fax +33 4 76 44 88 21; www-laog.obs.ujf-grenoble.fr

# 2. JSDC : CONSTRUCTION

## 2.1 Background

JMMC introduced the SearchCal web service to astronomers in 2004. It was built to be as dynamic as possible, in order to accommodate the rapidly evolving field of VO (Virtual Observatory) and mainstream optical interferometry. The goal was to build a scientific research tool able to query lots of different catalogs hosted at CDS to later compute properties from those data.

As time and knowledge evolved sufficiently to consider SearchCal computations reliable by field experts, new needs appeared, especially being able to retrieve stellar diameters without network connection. The VO architecture of SearchCal could not handle this specification. Thus came the idea to compile an all-sky catalog of diameters once and for all, by aggregating SearchCal results. This monolithic file could then be embeddable in any other software needing diameters, and would also be published to CDS.

## 2.2 Software architecture

To better understand the challenges and solutions undertaken to generate the JSDC, we first have to detail how the historic software works.

As shown in Figure 1, the system is a third-tier networked architecture, dispatched across the data (available at CDS), the scientific engine (hosted on JMMC servers), and the user front-end (dedicated to querying and result presentation).

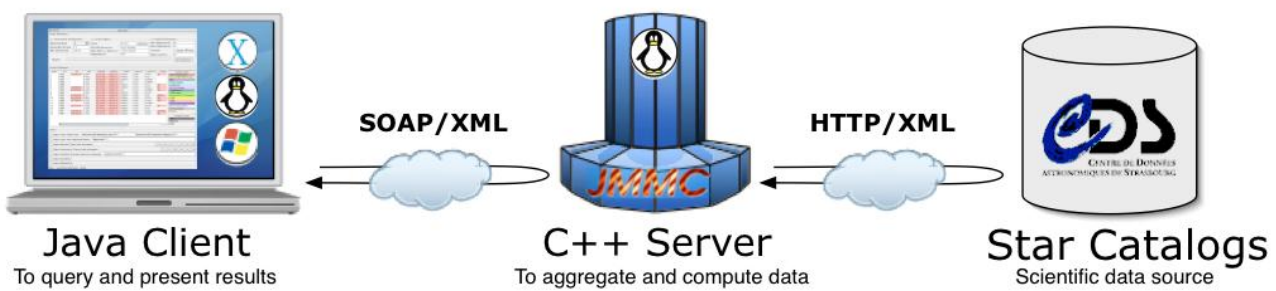

Figure 1. SearchCal software architecture.

The scientific core is on the server-side of the system, hosted at LAOG. This application is responsible for:

- Querying the CDS catalogs according to user-defined observational and instrumental configuration parameters (such as science object coordinates, search area, magnitude band and range, etc) received from the client application;
- Aggregating all those data in one summarizing table;
- Completing this table where data may be missing (R & I magnitudes, limb-darkened & uniform diameters, etc);
- Returning the final results to the client application for display and further filtering.

## 2.3 Scientific requirements

In order to create the largest catalog possible, we first had to define which scientific criteria this list should conform to. The JMMC Calibrator group specified the following constraints:

- The whole sky should be covered;
- The largest possible magnitude range in K band should be harvested, which goes from a theoretical -5.0 up to 20.0 for the current SearchCal engine;
- No inclusion of previously computed diameters extracted from foreign catalogs, to have a consistent computation method for each catalog's diameters;
- Tight result filtering, to eliminate obvious false positive like multiple stars of same color and so on.

- Catalog completeness, as no entry's property should be empty.
- Restricted column set, to clearly present diameters and star properties used to compute them.

**2.4 Technical solutions**

Once scientific goals were defined, we had to consider technical solutions that would fulfill those needs.

**2.4.1 Angular diameter determination**

For each star of a given box and magnitude range, SearchCal computes its angular diameter using a surface brightness method and calibrations for (B-V), (V-R) and (V-K) color indexes as presented in Delfosse & Bonneau[9]. Stars whose angular diameters estimated from the various color indexes are not comparable, are rejected, and a reliable error on the estimated diameter is computed (see Bonneau et al. 2006 for the details about rejection criterion and errors estimation). The LDD (Limb-Darkened angular Diameters) are obtained using the V magnitude and the (V-K) color index.

Then, values of the UDD (Uniform Disk angular Diameters) are computed for the photometric bands (B, V, R, I, J, H, K) and spectral types using the linear limb-darkened coefficients from Diaz-Cordoves J. et al.[10] and Claret A. et al[11].

**2.4.2 Sky mozaïcing**

SearchCal queries are bound to a maximum box size of 3600*1200 arcminutes. The first step to prepare our catalog build was to split the celestial sphere in discrete boxes to serialize querying. We wrote some *Python* scripts to handily generate batch querying configuration files for any given box size.

First attempts at creating the catalog shown some defects in sky coverage. The issue was the maximum number of stars the CDS can return for such a box in a given magnitude range. We had to find a compromise between boxes size, maximum number of returned stars, and query timeouts.

The good ratio was to ask for a thousand stars from -5.0 to 20.0 magnitude in a box of 180*120 arcminutes, for a total of 21600 queries. The final coverage is now uniform, as illustrated in Figure 2.

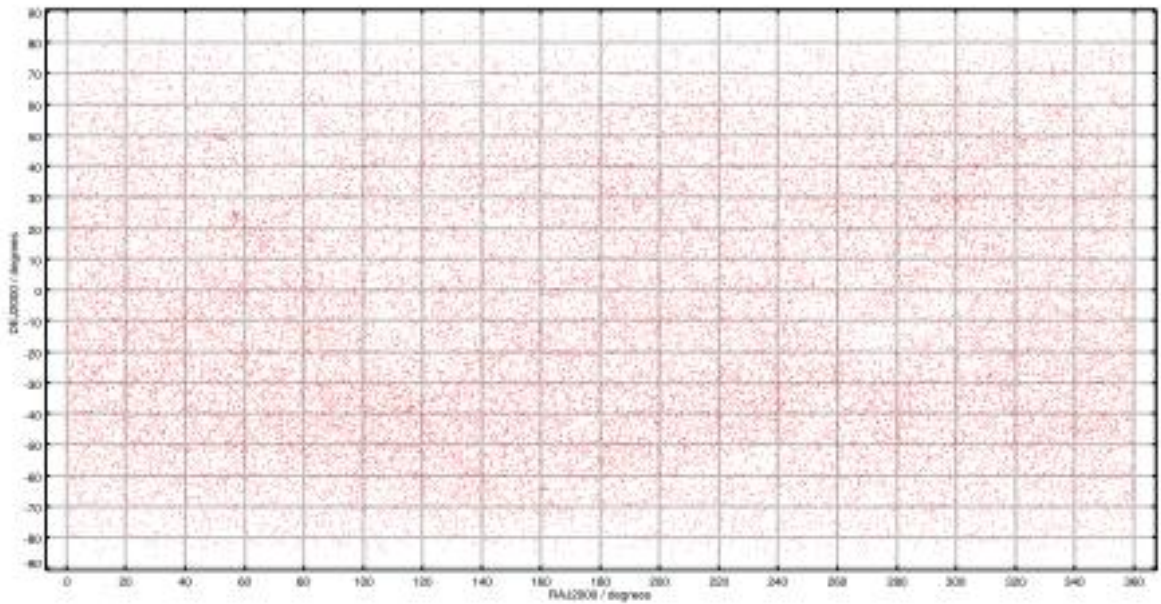

Figure 2. JSDC sky coverage.

### 2.4.3 Batch mode

We also had to be able to run queries by pools, automatically executing those 21600 requests one at a time. Fortunately, the SearchCal engine running on our server is also useable in batch mode, giving one result file in VOTable format for each request done. The time needed to generate one catalog is approximately 3 days.

### 2.4.4 Results aggregation

Each query gives us a VOTable file containing potential calibrators with computed diameters. The first step is to concatenate all those files into one big FITS file to speed up later computation and filtering. We made all the aggregation process with *bash* scripts using *XMLStarlet* to extract data from metadata in VOTables. Then we used *STILTS*[12] to convert the huge full catalog to FITS format.

### 2.4.5 Multiple instance filtering

Given that our search boxes may overlap, whose surface moreover varies as the cosine of their declination, we needed a robust algorithm to remove multiple instances of duplicated stars in the generated catalog. This problem was solved by extensively using *STILTS* again, to filter all stars with identical HIP, HD or DM id, or RA/DEC coordinates.

### 2.4.6 Stellar multiplicity filtering

To avoid specific confusion problems due to stellar multiplicity, we also decided to remove all stars known as spectroscopic binaries or as close visual pairs (angular separation $\leq 2$ arcseconds). We used 9th Catalog of Spectroscopic Binary Orbits[13] and Washington Visual Double Star Catalog[14] querying to finely flag problematic stars out.

In doing so, we discovered some mistakes in the epoch of some coordinates provided by those catalogs, as some well known multiple stars were still to be found in JSDC after filtering. It forced us to circumvent this inconsistency by manually querying *Simbad*[15] on stars identifiers for each catalog entry, instead of cross-matching *Vizier*[16] results on their sole coordinates.

### 2.4.7 Catalog completeness and format

Once more, to ensure no data was missing, we used *STILTS* to remove any catalog entry with any properties left blank.

Furthermore, JSDC is formatted to a restricted columns subset of typical SearchCal results, mainly star coordinates, proper motions, magnitudes, limb-darkened and uniform diameters, parallaxes, spectral types (with derived gravities and effective temperatures).

## 2.5 Results analysis

Before publishing JSDC to CDS, we took special attention to thoroughly study our catalog through systematic statistical analysis of the main relevant scientific indicators.

### 2.5.1 Parallaxes range and accuracy

The relative error $e\_Plx / Plx$ is always below 25%, as required by SearchCal algorithm for a reliable correction of the interstellar absorption.

All star of the catalog have a measured parallax between 2.01 mas (milliarcseconds) for the farthest, up to 333.79 mas.

### 2.5.2 Magnitudes statistics

The ranges of the stellar magnitudes found in JSDC are given in the Table 1. Figure 3 shows the distribution of magnitudes in the K band. Those ranges of magnitude are coherent with the limitations of the HIPPARCOS catalog (up to V magnitude 12.4).

Table 1. Magnitude ranges for each band.

| Magnitude Band | Minimum Value | Maximum Value |
|---|---|---|
| B | 0.019 | 13.642 |
| V | 0.074 | 12.225 |
| R | -0.52 | 11.34 |
| I | -1.31 | 11.339 |
| J | -1.84 | 11.246 |
| H | -2.59 | 11.376 |
| K | -2.81 | 11.48 |

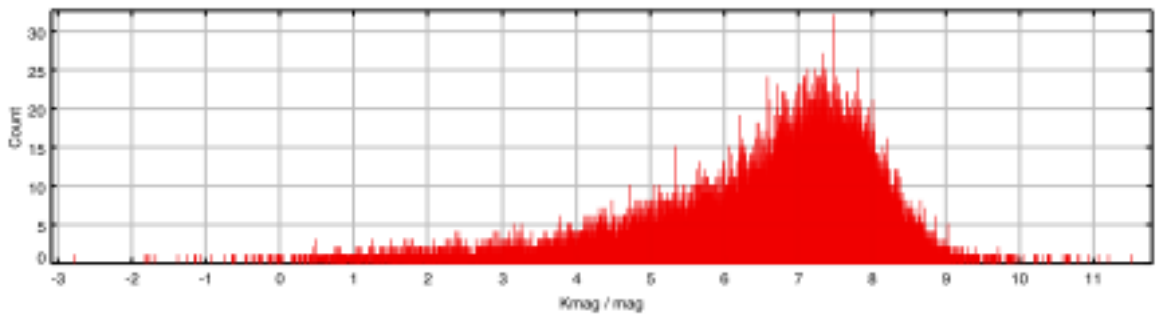

Figure 3. K magnitude distribution (steps of 5.0E-4).

### 2.5.3 Magnitude versus color diagrams

On Figure 4, we can see that the B-V and V-K diagrams all look fine. But for V-R and V-I some discretization is visible. By construction B, V and K-band magnitudes are extracted from catalogs of photometric measurements (only objects with existing measurements in this three band are retained). Therefore B-V and V-K color indexes take all possible values.

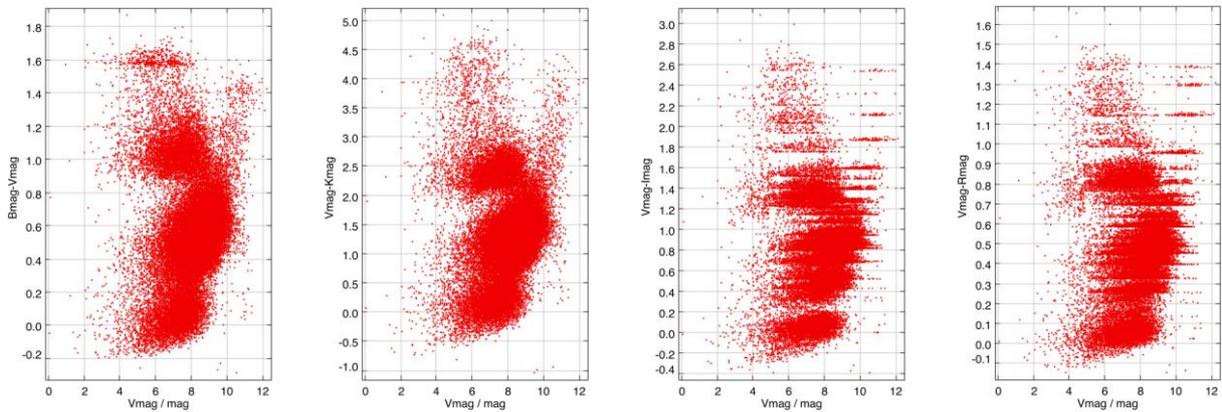

Figure 4. Magnitude vs. color diagrams.

On the other hand R and I-band magnitudes could be missing in the catalogs used. In such case they are computed from a spectral types-color relation (with an accuracy of approximately 0.1 magnitude). As spectral types have a digitalized input (the sub-spectral type), this involves discretization of V-R and V-I color indexes.

**2.5.4 HR diagram analysis**

Here we produce the HR diagram of our catalog in Figure 5. Please note that it was not generated using interstellar absorption-corrected magnitudes (not present in JSDC), but this should not make significant impact, as the majority of our sources are not that reddened anyway.

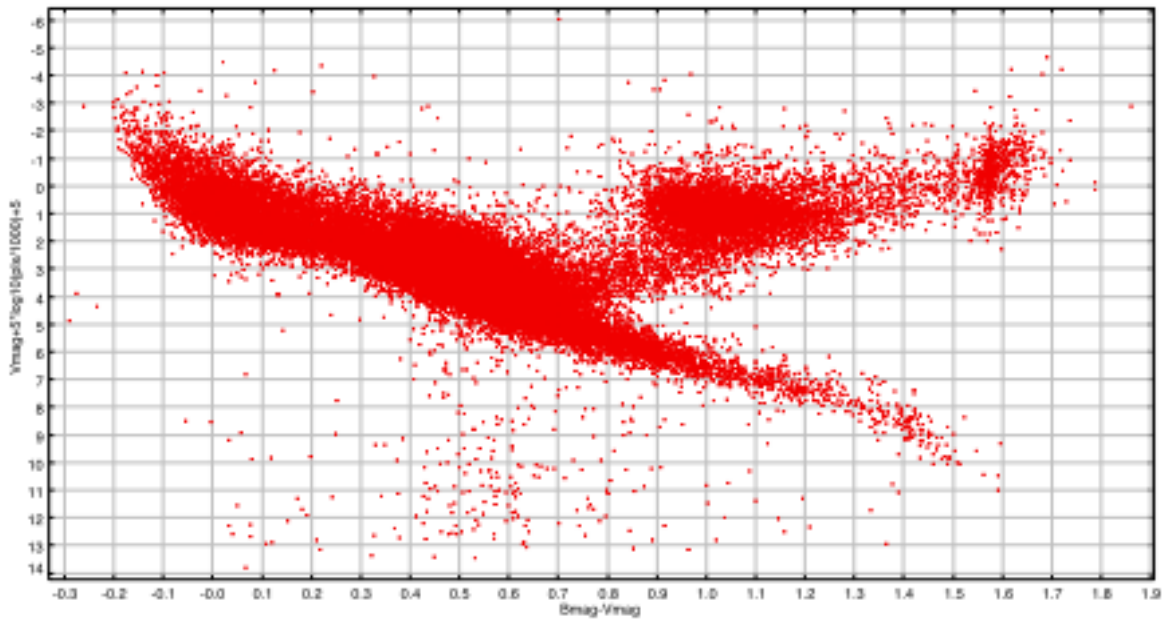

Figure 5. HR diagram.

The main sequence and giant sequence are clearly visible. For stars bluer than B-V=0.8 (O to G stars) the JSDC essentially contains main sequence objects. On the contrary, for the red part of the sample (B-V>0.8; K and M stars), the giants dominate.

**2.5.5 Computed diameters statistics**

JSDC LDD range from 0.01 mas up to 20.868 mas. LDD distribution is shown in Figure 6.

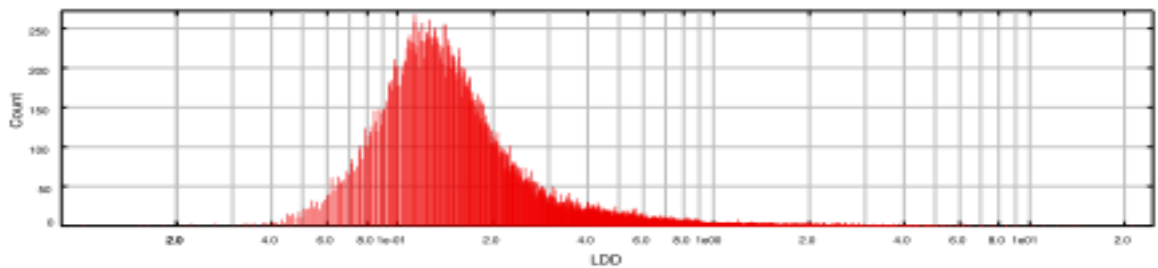

Figure 6. LDD logarithmic distribution.

UDD statistics go as follows in Table 2:

Table 2. UDD ranges, in mas.

| UDD Band | Minimum Value | Maximum Value |
|---|---|---|
| B | 0.009 | 18.842 |
| V | 0.009 | 19.054 |
| R | 0.01 | 19.36 |
| I | 0.01 | 19.668 |
| J | 0.01 | 19.974 |
| H | 0.01 | 20.136 |
| K | 0.01 | 20.273 |

The LDD distribution of JSDC stars in function of V-band and K-band apparent magnitude can be seen in Figure 7 and Figure 8 respectively.

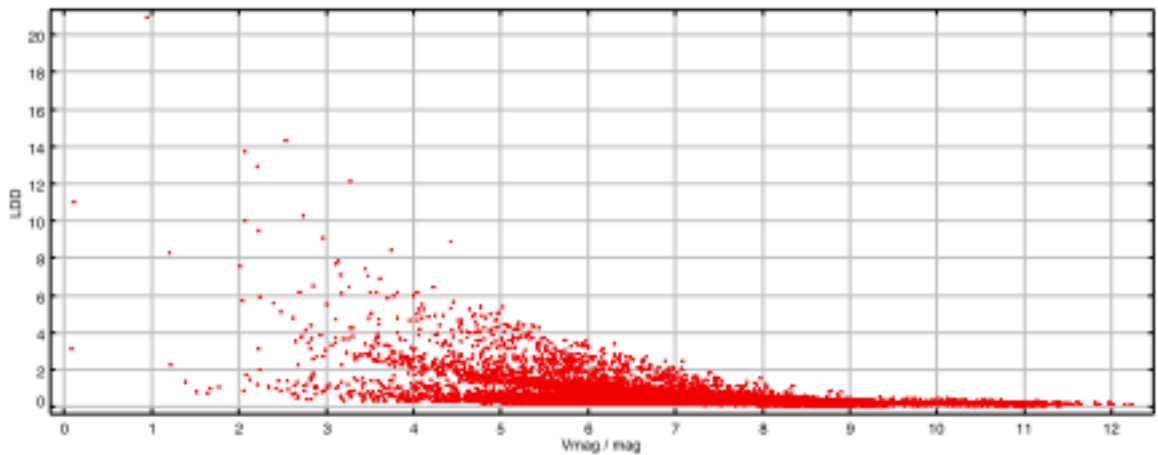

Figure 7. LDD vs. V-band diagram.

For a given V magnitude, LDD can take a wide range of value with an upper limit increasing with brightness.

But most remarkable is the behavior of LDD in function of K-band apparent magnitude. By construction, objects brighter than K=4 are mainly red giants stars. They have an effective temperature similar to each other, and their angular diameter depend mainly to apparent magnitude. So we bring up that, for bright objects, K magnitude gives a good estimation of the angular diameter.

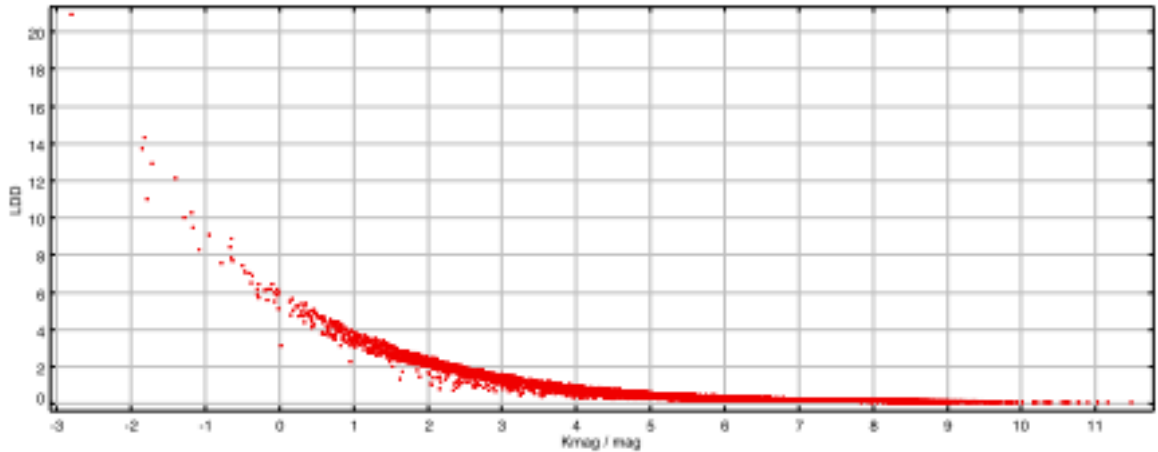

Figure 8. LDD vs. K-band diagram.

**2.5.6 Computed diameters compared to measured ones**

To check the validity of SearchCal statistical estimation of diameters, we compared our results with the few measured ones from the CHARM2 catalog[17] to estimate the diameters determination accuracy in JSDC. CHARM2 is a compilation of direct observation as well as indirect estimates of stellar diameters. For our comparison we only considered direct UDD measurements in K-band from long baseline interferometry, speckle or aperture masking. 82 stars are common to both catalogs with those constraints.

Figure 9 shows the relative difference (in %) between JSDC angular diameter estimation, and measured diameters in CHARM2.

A significant deviation exists for the lower diameters (<1.5 mas). We attribute this feature to bias in the measurements since there is no reason the estimated diameters should be less accurate for small radii than larger ones (as opposed to measures which are degraded). If we only consider angular diameter greater than 1.5 mas, the difference between mean measurements (CHARM2) and mean estimation (JSDC) is inferior to 1%. The RMS of the difference is 11%, much of it coming from measurement errors.

We confirm that the determinations of diameters in JSDC are not biased and that the diameters accuracy is better than 10%.

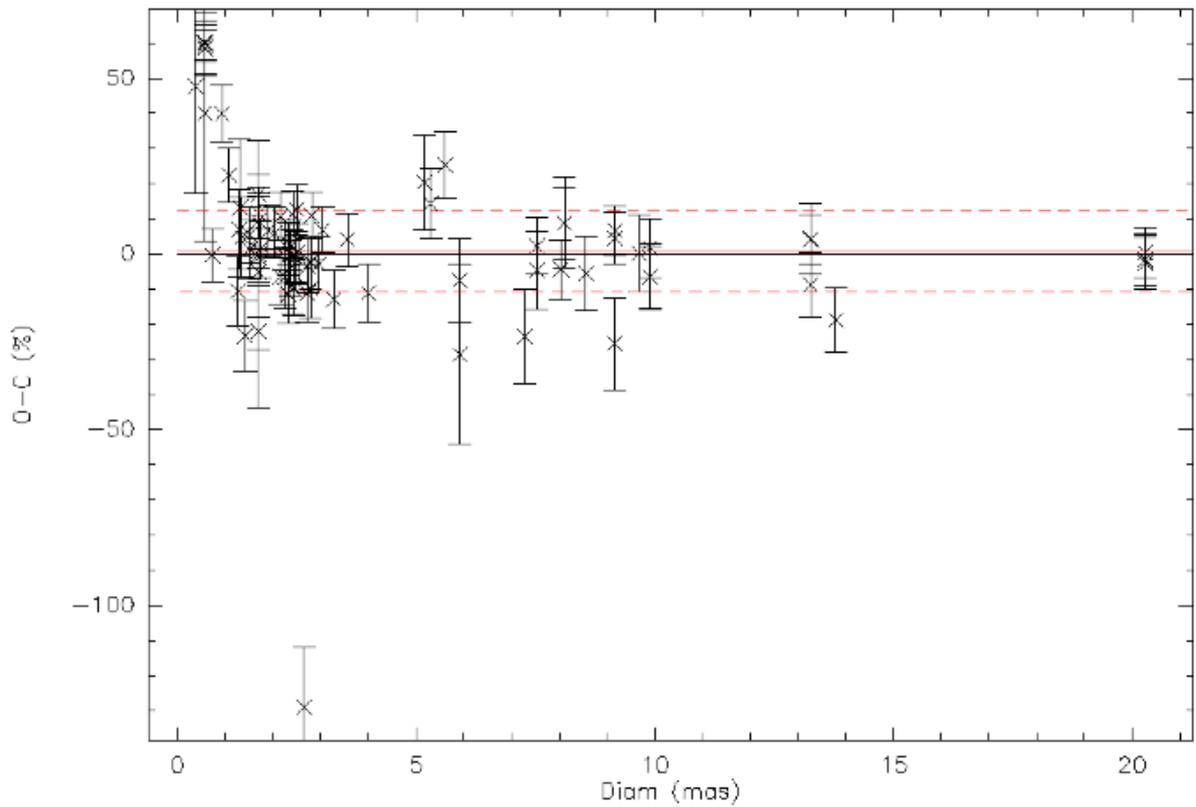

Figure 9. Relative difference between measured UDD (in K) in CHARM2 and determined one in JSDC. The full line is the mean value of the difference (0.7%) and the two dashed lines are a 1-sigma envelope (11.3%).

### 2.6 Conclusion

This large-scale study was very insightful, mainly to track very specific issues in SearchCal only detectable on large data sets rather than on selected particular test cases (most of them being tracked down to inconsistencies in several stellar catalogs).

One caveat was the lack of large measured diameters catalogs to better compare our statistic results against observed ones. But the comparison done shows confidence in the catalog we built.

JSDC is the largest diameter catalog available today, with more than 38 000 limb-darkened and uniform stellar diameters. It is on its way to CDS for publication, and will be further described and updated on http://www.jmmc.fr/jsdc

## 3. BADCAL

### 3.1 Background

IAU Commission 54 about Optical & Infrared Interferometry[18] has a dedicated Calibrator Stars workgroup that provides, among other things, a repository[19] to collect all observed stars that could not be used as interferometric calibrators.

In order to integrate this useful feedback in the results provided by SearchCal and thus JSDC, we needed a VO-compliant web service able to harvest the bad calibrator list.

JMMC undertook this effort, internally known as BadCal.

### 3.2 Scientific requirements

JMMC Calibrator Group defined the following needs for a new tool dedicated to bad calibrator management. The system should:

- Be synchronized with all known bad calibrators (i.e. IAU list at current time);
- Provide at least the same data set (submitter name and affiliation, justification and observational instrumental configuration, moderator comments…) as the original IAU list;
- Allow easy submission (using star name to automatically resolve RA/DEC coordinates from *Simbad*) of new suspicious calibrators;
- Ensure that calibrator submissions are moderated by at least one OLBI specialist. Each entry should come with observational instrumental configuration, plus scientific justification of its submission.
- Ascertain that the submitted calibrator name is not a known multiple star, but the name of the multiple system sub-star;
- Be interoperable with human and machine usage (i.e. VO-compliant querying interface, as cone search);
- Allow removal of false positive bad calibrators from the list while keeping history of each record. The moderator should afterward be able to confirm to users that a star was indeed present in the list previously;
- Provide entry removal only to its submitter or BadCal moderators. Contestants should first address their opposition to the provided submitter's email;

### 3.3 Technical solutions

BadCal is a *Java* Web application running on top of a *TOMCAT* server. This software handles the forms that manage bad calibrator submission, and displays the list of searched stars in synthetic or detailed *HTML* tables. The records are stored in a *MySQL* database. The search engine is delegated to *DSA*[20], a VO-compliant Web application developed by the Astrogrid community to serve astronomical catalogs in RDBMs. The user authentication is performed by one central accounting system hosted on the JMMC servers.

The combination makes it really easy to query one part of the sky given some coordinates and a search radius using the Web interface or scripts. Results are either exported as HTML tables, VOTables, or CSV formats. The SearchCal server will use this new service as soon as a *DSA* upgrade (which we are working on) provides VOTables in the ASTRORES style.

### 3.4 Conclusion

An advanced prototype of BadCal is currently ready at the JMMC, shared among beta-testers. It should go public quickly on http://www.jmmc.fr/badcal/

The main defaults of previous lists were that they were based on the good willing of observers, without help and automatic checks that help to rapidly isolate the calibrator as bad. With the multiplication of the facilities, now performing a wealth of observations with various instruments, each one providing many different observational characteristics, there is a new need for a wide communication of the bad calibrators corresponding to precise observing conditions.

BadCal is a flexible and versatile answer to this need, and should be of interest for the broad optical interferometry community. As such, we advocate a large advertising to its use.

### ACKNOWLEDGMENTS

This research has made use of the Jean-Marie Mariotti Center SearchCal service available at http://www.jmmc.fr/searchcal, co-developed by FIZEAU and LAOG.

This research has made use of the CDS Astronomical Databases SIMBAD and VIZIER available at http://cdsweb.u-strasbg.fr

This research has made use of the TOPCAT http://www.starlink.ac.uk/topcat/ and STILTS http://www.starlink.ac.uk/stilts/ software, provided by Mark Taylor of Bristol University, England.